\documentclass[aip,reprint]{revtex4-1}

\usepackage{amsmath}
\usepackage{graphicx}
\usepackage[dvips]{color}

\draft 

\begin{document}

\title{Alteration of helical vortex core without change in flow topology}

\author{Clara M. Velte}
\email[C.M. Velte: ]{cve@mek.dtu.dk}
\affiliation{Technical University of Denmark, Nils Koppels All\'{e} B403, 2800 Kgs Lyngby, Denmark}
\author{Valery L. Okulov}\affiliation{Technical University of Denmark, Nils Koppels All\'{e} B403, 2800 Kgs Lyngby, Denmark}
\author{Martin O.L. Hansen}\affiliation{Technical University of Denmark, Nils Koppels All\'{e} B403, 2800 Kgs Lyngby, Denmark}

\date{\today}

\begin{abstract}
The abrupt expansion of the slender vortex core with changes in flow topology is commonly known as vortex breakdown. We present new experimental observations of an alteration of the helical vortex core in wall bounded turbulent flow with abrupt growth in core size, but without change in flow topology. The helical symmetry as such is preserved, though the characteristic parameters of helical symmetry of the vortex core transfer from a smooth linear variation to a different trend under the influence of a non-uniform pressure gradient, causing an increase in helical pitch without changing its sign.
\end{abstract}


\maketitle

Vortex Generators (VGs), as first introduced by Taylor~\cite{Taylor}, are series of small winglets that are glued on a surface usually shortly upstream of an otherwise separated region. They are similar to wings with a small aspect ratio and are mounted normally to a surface with an angle of incidence to the oncoming flow. By creating streamwise vortices, they mix high momentum free stream fluid into the near-wall region of the boundary layer, thus delaying separation.

Previous investigations have shown that vortices generated by rectangular vanes on a flat plate have helical structure of the vortex lines and can thus be described by a simple vortex model~\cite{VelteJFM}. This model is characterized by a zero radial vorticity component, a fixed linear correlation between the axial and circumferential vorticity components and a Gaussian axial distribution~\cite{okulovetal}:
\begin{equation}\label{eqn:1}
\omega_r = 0;\quad \omega_{\theta} = r\omega_z/l;\quad \omega_z = \frac{\Gamma}{\pi \, \varepsilon^2}\exp{\left ( -\frac{r^2}{\varepsilon^2} \right )}
\end{equation}
where $\varepsilon$ is the radius of the vortex core, $\Gamma$ is the flow circulation, $2\pi l$ is the helical pitch, and the vorticity vector ($\omega_r$,$\omega_{\theta}$,$\omega_z$) is always directed along the tangent of the helical lines $x=r\cos (\theta)$; $y=r\sin (\theta)$; $z=l\theta$. The swirling flow generated by this vorticity field has a constant vortex advection velocity $u_0$ and can be described in terms of the local velocity components:
\begin{equation}\label{eqn:2}
\begin{split}
u_{\theta}& = \frac{\Gamma}{2\pi r}\left [ 1 - \exp \left ( -\frac{r^2}{\varepsilon^2} \right ) \right ]; \\
u_z &= u_0 - \frac{\Gamma}{2\pi l}\left [ 1 - \exp \left ( -\frac{r^2}{\varepsilon^2} \right ) \right ]; \quad \mathrm{or}
\end{split}
\end{equation}
The relation between velocity components $u_z$ and $u_{\theta}$ provides a simple condition for helical symmetry~\cite{okulovetal} (the velocity formulation):
\begin{equation}\label{eqn:2b}
u_z = u_0 - u_{\theta}r/l
\end{equation}
Note that the same vortex model (\ref{eqn:2}), in the sense of the Batchelor's or q-vortex~\cite{VelteJFM,okulovetal}, is commonly used in instability studies of swirling flows to describe vortex breakdown~\cite{okulovetal,Leibovich,Escudier,HeatonPeake}. This well-known phenomenon is characterized by three specific features~\cite{Leibovich}: (1) a change in flow topology; (2) a change in axial velocity from a jet-like profile before breakdown to a wake-like profile after breakdown; (3) a structural change resulting in vortex core expansion. In addition, model (1) together with a simple control volume (CV) analysis has been successfully applied~\cite{Okulov1996,OkulovSoerensen2010} to predict the velocity profiles after breakdown in swirling pipe flows. The results of the CV analysis are able to explain the growth of the vortex core behind vortex breakdown that is seen in the experimental data. Moreover, they support the hypothesis that the change in axial velocity from a jet-like profile to a wake-like one during vortex breakdown should be associated with a transition in helical symmetry of the vortex structure, which was simulated in Okulov \textit{et al.}~\cite{Okulov}.

Even more interestingly, however, the results of this CV-analysis predict an atypical transition where the changes in the helical parameters of a vortex (\ref{eqn:1}) may appear without accompanying changes in the flow topology and transition from a jet-type axial velocity profile to a wake-type one. Consider examples from Okulov and S{\o}rensen~\cite{OkulovSoerensen2010} and Martemianov and Okulov~\cite{Martemianov} where up to 4 types of axisymmetrical helical vortices were analytically found under the same fixed integral flow characteristics (flow rate; velocity circulation; axial flux of angular momentum; axial flux of momentum; and axial flux of energy) corresponding to a concrete operating regime of the pipe swirling flow in the experiments. The first vortex from the solutions has right-handed helical structure of the vortex lines, yielding a jet-like axial velocity profile, and the remaining three vortices have left-handed helical symmetry with wake-like profiles. Clearly, the flow may be subject to transitions between all the different vortices which may exist under the same integral flow characteristics. So transition should be possible between the jet- and wake-like profiles (typical vortex breakdown state) as well as between the different solutions for wake-like profiles (atypical transition). With this previous work in mind, the current study addresses experimental observations of this unknown second type of vortex transition.

\begin{figure}
\includegraphics[width=1.0\linewidth]{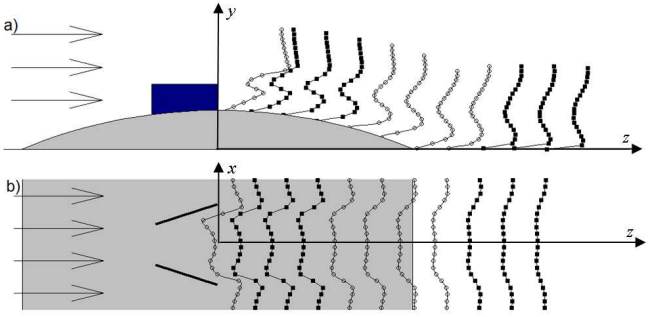}
\caption{Vertical (a) and horizontal (b) cross-section of the testing flow with axial and vertical velocity profiles, respectively: lines with circles indicate zones where the vortex changes (near wake and the zone of the vortex transition); lines with filled squares indicate zones with stable vortex.\label{fig:1}}
\end{figure}

\begin{figure}
\includegraphics[width=1.0\linewidth]{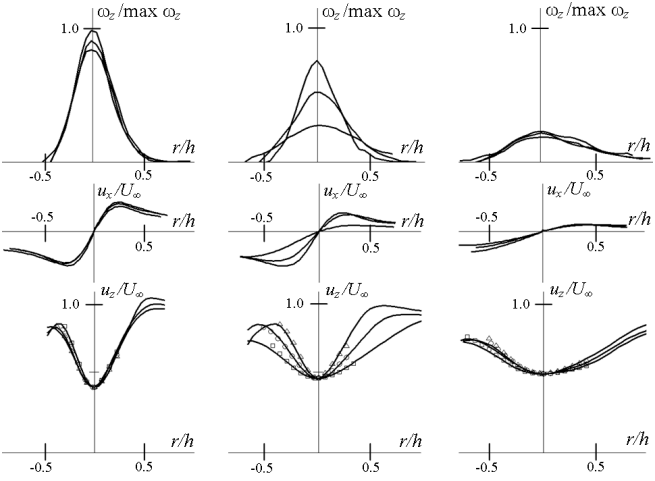}
\caption{Profiles of the axial vorticity (top row); azimuthal (middle row) and axial (bottom row) velocity in local coordinates for different vortex states of the far wake: the left and right columns coincide with the two stable zones $1< z/h <4$ and $7< z/h <10$ where the profiles persist without big changes; the middle column represents $4< z/h <7$ where a strong change in the vortex core was detected. The vertical vorticity and velocity profiles were extracted from the averaged stereoscopic PIV velocity field through the center of the vortex core. The symbols in the lower row display profiles recalculated by the helical symmetry condition (\ref{eqn:2b}).\label{fig:22}}
\end{figure}

Measurements have been conducted in wall bounded flow in a wind tunnel on a cylindrical sector (bump) downstream of a vortex generator cascade (see Figures~\ref{fig:1}\,\textit{a} and \textit{b}) at a low Reynolds number (free stream velocity U$_{\infty}=1.0$\,ms$^{-1}$). The closed-circuit wind tunnel has a 8:1 contraction ratio and a test section of cross-sectional area 300$\times$600\,mm with length 2\,m. The sector had a radius of 390\,mm, a span of 600\,mm, a width of 300\,mm and a height of 30\,mm. The coordinate system is defined at each measurement volume in Figure~\ref{fig:1}. z is the axial flow direction, y the wall-normal and x the spanwise one. The vanes were placed in a cascade along the span of the bump (see Velte~\cite{Velte} for exact configuration), in a fashion inducing counter rotating vortices, with the trailing edges at the bump center. The vane height was chosen based on the structure of the boundary layer, i.e., they have the same height.

The flow was measured in 14 spanwise planes (x-y), not all of which are shown in Figure~\ref{fig:1}, in a range of streamwise positions z. The stereoscopic PIV equipment included a double cavity NewWave Solo 120XT Nd-YAG laser (wavelength 532\,nm), capable of delivering light pulses of 120\,mJ. The light sheet thickness at the measurement position was 2\,mm. The equipment also included two Dantec Dynamics HiSense MkII cameras (1344$\times$1024\,pixels) equipped with 60\,mm lenses. Both cameras were mounted on Scheimpflug angle adjustable mountings. The seeding, which was added downstream of the test section in the closed-loop system, consisted of DEHS droplets with a diameter of 2-3\,$\mu$m. The laser was placed above the test section, illuminating a plane normal to the test section walls. One camera was placed in the forward scattering direction and the other one in the backward scattering one. The angle of each respective camera to the laser sheet was 45$^{\circ}$. The f-numbers of the cameras were set to between 8 and 16 (forward scattering) and 4 and 5.6 (backward scattering). The SPIV setup was rigidly mounted on a traverse which could translate the setup along the test section in the streamwise direction. In this manner, the calibration needed to be performed only once. The images were processed using Dantec DynamicStudio software version 2.0. For each measurement position, 500 independent realizations were acquired. The disparity due to misalignment between the centre of the light sheet and the calibration target found in these investigations, typically around 0.05 pixels, were always smaller than the optimal measurement accuracy of the PIV system ($\sim$0.1 pixels).

As an overview of the development of the vortices behind a cascade of vortex generators, Figure~\ref{fig:1} displays vertical and horizontal cross-sections of the test region with axial and vertical (azimuthal) velocity profiles in their positions of acquisition along the bump and wind tunnel test section wall, respectively. From the velocity plots it is clear to see that each generator in the cascade induces only one strong vortex, but not the accompanying secondary vortex which was well pronounced in our former experiments with a single vortex generator~\cite{VelteJFM}. The absence of these weaker secondary structures in the present case simplified our study, reducing the result of each vortex generator to merely one primary vortex. The wake behind vortex generators is usually divided into three parts: near wake portrayed by the first profile in Figure~\ref{fig:1} ($z/h<1$) where the vortex forms; far wake ($1< z/h <10$ and above) where the vortex is well pronounced; and lastly a zone behind the far wake where the vortex core can not be separated from other flow disturbances. The current study focuses on the far wake. The testing regime in Figure~\ref{fig:1} can be separated into 3 zones of the vortex development in the far wake: the first one is $1< z/h <4$ where the vortex persists without big changes; the second one is $4< z/h <7$ where a strong change in the vortex core is detected; and the last one is $7< z/h <10$ where the vortex core returns to a stable state. To make this result more clear, we have appended an analysis of the local velocity and vorticity fields around the vortex core for all cross-section in Figure~\ref{fig:22}. The profiles are extracted at vertical lines through the vortex centers. The slight asymmetry of the profiles appear because the vortex is embedded in the non-uniform flow of the boundary layer. We have grouped the profiles into three different stages in accordance with our visual division of the vortex development. Comparisons of the velocity profiles proved that zones 1 and 3 coincide with stable vortex stages and zone 2 describes a strong transition in the vortex core.

\begin{figure}
\includegraphics[width=1.0\linewidth]{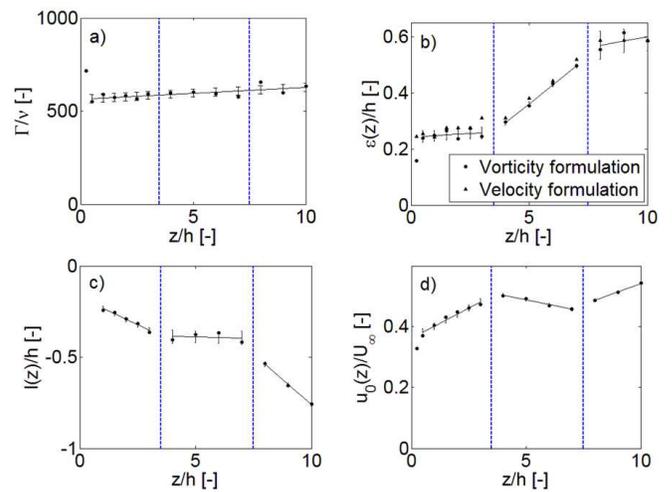}
\caption{Helical parameters of the actuator induced vortex. The parameters have been non-dimensionalized by the vortex generator height $h$, viscosity $\nu$ and free stream velocity U$_{\infty}$, respectively. The parameters presented in each subfigure are (a) estimate of the circulation, (b) the vortex core radius, (c) helical pitch and (d) vortex convection velocity.\label{fig:4}}
\end{figure}

The linear velocity formulation for helical symmetry (\ref{eqn:2b}) allows the usage of averaged velocity fields for determination of the helical vortex characteristics from of the turbulent flow. We can therefore neglect some small meandering of the vortex core and reflect the behavior of the vortex parameters of the fitting model (\ref{eqn:2}) to the averaged flow characteristics as a first approximation of the flow. Figures~\ref{fig:4}\,\textit{a-d} show the helical fitting parameters of (\ref{eqn:2}) as a function of the downstream distance from the actuator cascade. The quantities are non-dimensionalized by the vortex generator height $h$, free stream velocity U$_{\infty}$ and kinematic air viscosity $\nu$. The circulation $\Gamma$ and the vortex core radius $\varepsilon$ were obtained from fitting the expression for the Gaussian vorticity distribution (\ref{eqn:2}) at a number of azimuthal positions after imposing a polar grid on the Cartesian one and choosing origo at the vortex center. It is important to note that this value of the circulation may not be the same as that of the real unsteady flow, since it describes the circulation of the stationary vortex model (\ref{eqn:2}) applied in Velte \textit{et al.}~\cite{VelteJFM} to fit the averaged flow field. Hence, this model does not account for possible meandering effects of the vortex core in the turbulent boundary layer. Nevertheless, as is expected from the flow, the circulation of our model does not change considerably along the downstream evolution of the vortex except for the variations when the vortex is initially formed (Figure~\ref{fig:4}\,\textit{a}). In addition to the estimate obtained from fitting the model to the data, the vortex core radius was estimated by direct measurement of the position of maximum azimuthal velocity in local coordinates. In Figure~\ref{fig:4}\,\textit{b}, the two estimates show good agreement, indicating a small change of the vortex core in the first and third stable zones and a growth of the vortex core of a factor of about two across the second zone.

As for the remaining parameters, the convection velocity $u_0$, being the axial velocity at the center of the vortex core, was obtained directly from the measurement data and the helical pitch, $2\pi l$, was obtained by minimizing the residual $0=u_0-u_r/l-u_z$ in a least squares sense. The resulting streamwise velocity profiles obtained from (\ref{eqn:2b}) using these fitting parameters are indicated by symbols in the lower row of Figure~\ref{fig:22}. The collapse of these points with the original axial velocity profiles confirm the simple condition for helical symmetry (\ref{eqn:2b}) of the vortex core. The downstream variation of these parameters (Figures~\ref{fig:4}\,\textit{c-d}) provides further support that the vortex in the far wake evolves through three zones where it can be approximated to vary linearly in the downstream direction but with different angles of inclinations of the lines in each zone. It should be noted that the trailing edge of the bump ($z/h =6$), where the geometry experiences a discontinuity from the cylindrical sector to the flat wind tunnel wall, does not significantly affect the vortex transition in the second zone ($4< z/h <7$). This is evident by the linear approximations in a least squares sense of the data with corresponding error bars, see Figures~\ref{fig:4}\,\textit{c-d}.

The pressure gradient, growing along the bump, triggers the abrupt transition to a new vortex state causing new behaviors of the flow parameters, which digress from the previous trend. In contrast to the typical transition, vortex breakdown with different orientation in symmetry of the vortex lines before and after transition from jet to wake, the new transition should be identified as an atypical form, since the sign of the negative helical pitch is preserved (Figure~\ref{fig:4}\,\textit{c}). In other words the longitudinal (axial) velocity profiles have the same wake-like forms with minimum velocity on the vortex axis (indicated in Figure~\ref{fig:1}\,\textit{a} and the lower row of Figure~\ref{fig:22}) with the same orientation of symmetry of the vortex lines (left to left transition) across the far wake. Further properties of this new phenomenon are similar to the classical vortex breakdown. Indeed this atypical transition takes place under the influence of a slight variation in the pressure gradient along the bump section. Further, the vortex core grows with a factor of about two behind this transition (Figure~\ref{fig:4}\,\textit{b}). These distinctions and sameness of the both transitions were found also in the CV analysis and demonstrated in Table 2 of Martemianov and Okulov~\cite{Martemianov}. Unfortunately until now, it seems that the lighter transformation of the vortex during the second type of transition has not received proper attention among investigators.

Our results show that the cascade of vortex generators produce longitudinal vortices that possess helical symmetry, as observed previously for a single vane in zero pressure gradient flow~\cite{VelteJFM}. The effects of the secondary vorticity are, however, substantially reduced by mounting the generators in a cascade in a fashion producing counter-rotating primary vortices that are spaced densely enough. Hence, the net effect of each actuator is basically a monopole. As expected, the circulation does not change much in the downstream direction. The remaining helical parameters display a smooth, linear variation along the three zones in the far wake, developing first through a stable vortex state through a transition zone and then back to a stable state. This observation supports the existence of a weaker second type of transition in the vortex core, where the sign of the helical symmetry of the vortex lines (with the same wake-like axial vortex profiles) and flow topology remains the same. Explanations to these observations can be found in previous theoretical work~\cite{Martemianov,OkulovSoerensen2010}. Furthermore, results of numerical simulations~\cite{Okulov} support these new experimental data. Note that this lighter transition between vortices of the same orientation of helical symmetry is subject to less dramatic changes in the vortex flow and does usually not display the changes in flow topology that have previously been observed during vortex breakdown.

\begin{acknowledgments}
This work formed a portion of the Ph.D. dissertation of CMV, supported by the Danish Research Council, DSF, under grant number 2104-04-0020, and currently by EUDP-2009-II-grant Journal number 64009-0279, which are both gratefully acknowledged.
\end{acknowledgments}

\providecommand{\noopsort}[1]{}\providecommand{\singleletter}[1]{#1}%

\end{document}